\documentclass[twocolumn,english,aps,superscriptaddress, pra,groupedaddress]{revtex4-1}

\usepackage[latin9]{inputenc}
\usepackage{amsmath}
\usepackage{amssymb}
\usepackage{graphicx}
\usepackage{babel}
\usepackage{mathrsfs}
\usepackage{amsfonts}
\usepackage{epstopdf}
\usepackage{caption}
\usepackage{multirow}

\begin{document}

\title{Virial Expansion for a Three-Component Fermi Gas in One Dimension: \\ The Quantum Anomaly Correspondence}

\author{Jeff Maki$^1$}
\author{Carlos R. Ord\'o\~nez$^{2,3}$}

\affiliation{1) Department of Physics and Center of Theoretical and Computational Physics, University of Hong Kong, Hong Kong, China \\ 2) Physics Department, University of Houston, Houston, Texas, 77024-5005, USA \\
3) ICAB, Universidad de Panama, Panama, Republica de Panama}

\date{\today}

\begin{abstract} 
In this paper we explore the transport properties of three-component Fermi gases confined to one spatial dimension, interacting via a three-body interaction, in the high temperature limit. At the classical level, the three-body interaction is scale invariant in one dimension. However, upon quantization, an anomaly appears which breaks the scale invariance. This is very similar to the physics of two-component fermions in two spatial dimensions, where the two-body interaction is also anomalous. Previous studies have already hinted that the physics of these two systems are intimately related. Here we expand upon those studies by examining the thermodynamic properties of this anomalous one dimensional system in the high temperature limit. We show there is an exact mapping between the traditional two-body anomalous interaction in two  dimensions, to that of three-body interaction in one dimension. This result is valid in the high temperature limit, where the thermodynamics can be understood in terms of few-body correlations.
\end{abstract}

\maketitle

\section{Introduction}

Symmetry is an important tool in understanding any physical system. For this reason, it is not surprising that when a classical symmetry is unexpectedly broken upon quantization, a phenomenon known as a quantum anomaly, it can create quite a stir among physicits \cite{Jackiw}. In cold atom experiments, one such anomaly to be predicted and observed was the scale anomaly in two dimensional Fermi gases \cite{Olshanii10,Hofmann12,Gao12}.

The two dimensional Fermi gas with short ranged two-body interactions (henceforth simply called the anomalous two dimensional Fermi gas) is classically scale invariant \cite{Rosch97}. If this symmetry were present under quantization, it  would drastically reduce the complexity of the energetics and dynamics \cite{Rosch97,Chin11, Kohl12, Dalibard14, Maki19}. However, this is not truly the case here upon quantization, as the two-body potential can support a bound state \cite{Olshanii10, Ordonez18b}. The presence of this new energy scale explicitly breaks the scale symmetry of the classical model. In this case, the breaking of scale invariance is logarithmically weak, and there have been numerous theoretical and experimental studies examining to what extent scale symmetry and the quantum  anomaly are present in the physics of two dimensional quantum gases \cite{Olshanii10,Hofmann12,Gao12,  Kohl12, Enss19, Chin11, Dalibard11, Dalibard14}.

Recently, a number of additional anomalous systems have been identified: bosons with three-body interactions in one spatial dimension \cite{Nishida18}, three-component fermions with three-body interactions in one spatial dimension \cite{Ordonez18a}, and the one dimensional quantum gas with a derivative coupling \cite{Camblong19}. These systems, which we will simply call anomalous one dimensional quantum gases, are classically scale invariant, but upon quantization, a bound state appears. For the case of a one dimensional quantum gas with three-body interactions, previous studies have shown that the coupling constant varies logarithmically with the bound state energy - just as the two dimensional quantum anomaly. This result has been recently used to study a number of thermodynamic and dynamic properties of these one dimensional anomalous quantum gases \cite{Nishida18, Ordonez18a, Drut18, Valiente19, Valiente19b, Pastukhov19}. One particular facet of these systems was noted in Ref.~\cite{Ordonez18a}, namely the logarithmic breaking of scale invariance led to a mapping between the physics of two dimensional fermions and that of these anomalous three-component fermions in one dimension, which we call the anomaly correspondence. In particular it was shown that the third virial coefficient, $\delta b_3$ for the anomalous one dimensional Fermi gas is directly related to its two dimensional counterpart.

Our goal is to explicitly test this analogy by computing the thermodynamic and transport properties of these three-component fermions in the large temperature limit. First, it is necessary to check whether the thermodynamics of the system obey the anomaly correspondence. To check this we focus on the virial coefficients and Tan's contact \cite{Tan08}, which have been shown to be related to the two dimensional anomalous Fermi gas \cite{Ordonez18a}. Once the thermodynamic properties have been examined, we proceed to calculate the bulk viscosity.


Fundamentally speaking, scale invariant systems in the normal phase have a vanishing bulk viscosity \cite{Son07, Taylor10}. For this reason the bulk viscosity is an important quantity in understanding the breaking of scale invariance, whether it be explicit or anomalous. Although one is often concerned with the static bulk viscosity, it is useful to consider the spectral function of the bulk viscosity, $\zeta(\omega)$. This quantity has been calculated in the high temperature limit for fermions with two-body interactions in a variety of spatial dimensions \cite{Nishida19, Enss19, Schaefer13a, Schaefer13b, Hoffman19}. 

In order to calculate the thermodynamic properties of the one dimensional anomalous Fermi gas, we perform the virial expansion to third order in the fugacity, $z$, following the arguments presented in Ref.~\cite{Nishida19}. We explicitly calculate the virial coefficients, Tan's contact, and bulk viscosity for the one dimensional anomalous fermions, and show that they are indeed proportional to their two dimensional counterparts. This allows us to explicitly verify this mapping between anomalous systems, which was previously based on scaling arguments \cite{Ordonez18a}, and to construct a dictionary for the anomaly correspondence.

The remainder of the article is organized as follows: in section \ref{sec:review} we review the few body physics of both the two-dimensional and one-dimensional anomalous fermions. We then apply this approach to calculate the shift in the third virial coefficient and Tan's contact in section \ref{sec:p_and_C}. In section \ref{sec:bulk_visc} we then compute the bulk viscosity. We then conclude in section \ref{sec:conc}.

\section{Review of the Few-Body Physics}
\label{sec:review}
We begin by reviewing the few-body physics of the three-component anomalous fermions, and how it relates to the standard anomalous paradigm in two spatial dimensions. The Hamiltonian for the anomalous three-component fermions is:

\begin{align}
H &= \sum_{\sigma = 1}^{3}\sum_k \frac{k^2}{2}\psi_{\sigma}^{\dagger}(k)\psi_{\sigma}(k) \nonumber \\
&+ \frac{g}{L^2} \sum_{k_i, l_i} \psi_1^{\dagger}(k_1)\psi_2^{\dagger}(k_2)\psi_3^{\dagger}(k_3)\psi_3(l_3)\psi_2(l_2)\psi_1(l_1) \nonumber \\ & \indent \delta_{k_1+k_2+k_3, l_1+l_2+l_3},
\label{eq:Hamiltonian}
\end{align}

\noindent where $\psi_{\sigma}(k)$ is the field operator that annihilates a fermion with spin, $\sigma = 1,2,3$, and momenta $k$, while $L$ is the length of the system, and the sum is over all six momenta. Naively one would expect that $g$ is dimensionless, however, this model is ultra-violet (UV) divergent, and depends on a short distance cut-off, $\Lambda$. The act of removing this length scale from the problem, will produce the quantum anomaly. 

\begin{figure}
\includegraphics[scale=0.45]{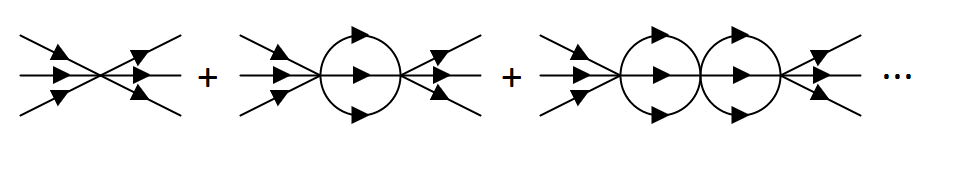}
\caption{Feynman diagrams that lead to the three body scattering amplitude, $T_3(Q,Q_0)$. Each line corresponds to a free fermionic propagator, while each vertex is a three body interaction, $g$.}
\label{fig:diagrams}
\end{figure}

To understand this UV divergence, consider the three-body scattering amplitude in the presence of the vacuum, $T_3(Q,Q_0)$, where the scattering amplitude is a function of the center of mass momentum, $Q$, and energy, $Q_0$. In this case the three-body scattering amplitude can be found as the summation of the diagrams shown in Fig.~\ref{fig:diagrams}. The result is:

\begin{align}
T_3^{-1}&(Q,Q_0) = \frac{1}{g}  \nonumber \\
&- \frac{2}{\sqrt{3}} \int_{-\infty}^{\infty} \frac{dp}{2\pi}\int_{-\infty}^{\infty} \frac{dq}{2\pi} \frac{1}{Q_0 -Q^2/6 - p^2 -q^2+i \delta},
\end{align}

\noindent  As one can see $T_3(Q,Q_0) = T_3(0,Q_0-Q^2/6)$, which is required by Galilean invariance. Therefore, we define the parameter $\epsilon = Q_0 - Q^2/6$, the energy of the relative motion of the three particles, and $T_3(Q,Q_0)$ will only be a function of $\epsilon$. Upon performing the integration over the intermediate momenta, one obtains:

\begin{equation}
T_3^{-1}(\epsilon + i \delta) = \frac{1}{g} + \frac{\ln\left( \frac{\Lambda^2}{-\epsilon -i \delta}\right)} {2 \pi \sqrt{3}}.
\end{equation}

In order to remove the UV dependence, we note that the physical bound state, $E = - E_B$, is defined as the pole of the three body scattering amplitude at zero center of mass momentum. This identification leads to the following expression for the coupling constant:

\begin{equation}
g = - \frac{2\pi \sqrt{3}}{\ln\left(\frac{\Lambda^2}{E_B}\right)}.
\label{eq:g_running}
\end{equation}

\noindent Eq.~(\ref{eq:g_running}) states the coupling constant is no longer a constant but a function of the new energy scale, the bound state energy, $-E_B$. Using Eq.~(\ref{eq:g_running}), we can eliminate the UV divergence, and express the T-matrix in terms of physical quantities:

\begin{equation}
T_3(\epsilon + i \delta) = \frac{2 \pi \sqrt{3}}{\ln\left(\frac{E_B}{-\epsilon - i \delta} \right)}.
\label{eq:t3}
\end{equation}

This should be compared to the two-body scattering amplitude:

\begin{equation}
T_2(\epsilon + i \delta) = \frac{4\pi}{\ln\left(\frac{E_{B}}{-\epsilon - i \delta}\right)},
\end{equation}

\noindent and the two-body coupling constant:

\begin{equation}
g_{2D} = - \frac{4\pi}{\ln\left( \frac{\Lambda^2}{E_{B}}\right)}.
\end{equation}

Assuming that the scattering properties of the two models can be matched, one can easily see:

\begin{equation}
g = \frac{\sqrt{3}}{2} g_{2D}.
\label{eq:g_ratio}
\end{equation}

In the following sections, we will exploit this fact to show that the connection between these two anomalous systems runs deeper, leading to a mapping between thermodynamic quantities in the high temperature limit.

\section{Calculation of the Pressure and Tan's Contact}
\label{sec:p_and_C}

In order to study the thermodynamic and transport properties of this anomalous system, we employ the virial expansion. For full details we refer the reader to Ref. \cite{Nishida19}. The main idea is to split the partition function into $N$-body sectors:

\begin{equation}
Z= \sum_{N=0}^{\infty} z^N Tr_N\left[ e^{-\beta H} \right],
\end{equation}

\noindent where $z = e^{\beta \mu}$ is the fugacity, $\beta = 1/T$, $\mu$ is the chemical potential, and $Tr_N$ denotes the trace over the $N$-body sector of the Hilbert space. 

In the high temperature limit, $z \ll 1$. This allows one to expand the partition function in terms of the fugacity, and to consider only a small number of few-body contributions to the partition function. For our purposes, we will work up to $O(z^3)$, or equivalently to $N=3$, as this is the first non-trivial order where interaction effects appear.

To evaluate the $N$-body partition function we need the matrix elements of the evolution operator $e^{-\beta H}$. In general this is an impossible task. However at the few body level, we can obtain an analytic result by employing the following identity:

\begin{equation}
e^{-\beta H} = \int_{-\infty}^{\infty}\frac{dE}{\pi} e^{-\beta E} Im \left[\frac{1}{E-H- i \delta}\right].
\label{eq:prop}
\end{equation}

\noindent At first and second order in the fugacity, the Hamiltonian is simply the non-interacting Hamiltonian, $H_0$. At $O(z^3)$ we will need to include the effect of interactions. The exact propagator at the three-body level can be evaluated \cite{Nishida19}, and is related to the three body scattering amplitude defined in Eq.~(\ref{eq:t3}):

\begin{align}
e^{-\beta H} &= \int_{-\infty}^{\infty}\frac{dE}{\pi} e^{-\beta E} Im \left[\frac{1}{E-H_0- i \delta} \right. \nonumber \\
& \left. +\frac{1}{E-H_0 - i \delta} T_3 \frac{1}{E-H_0 - i \delta}\right], 
\label{eq:prop_int}
\end{align}

\noindent where $T_3$ is the scattering amplitude operator which has matrix elements that only depend on the center of mass energy and momentum: $T_3(E-Q^2/6- i \delta)$.

The partition function, and subsequently the pressure, is evaluated by tracing  Eqs.~(\ref{eq:prop}) and (\ref{eq:prop_int}) over all three particle states, see Appendix \ref{app:z} for explicit expressions. From there, one can use the relationship between the partition function and the pressure:

\begin{equation}
\beta P L = \ln(Z),
\end{equation}

\noindent to obtain the following expression for the pressure:

\begin{align}
P &= \frac{3T}{\lambda_T}\left[z - \frac{1}{2 \sqrt{2}} z^2 + \frac{z^3}{3\sqrt{3}} \right. \nonumber \\
&\left. + \frac{z^3}{6\pi} \int_{-\infty}^{\infty}d \epsilon \ e^{-\beta \epsilon} Im \left[ \frac{T_3(\epsilon-i \delta)}{-\epsilon + i \delta}\right] \right].
\label{eq:pressure}
\end{align}

\noindent The first three terms of Eq.~(\ref{eq:pressure}) are the non-interacting contributions to the pressure, while the last term is due to the three body interactions. 

We now define the virial coefficients, $b_n$ via:

\begin{equation}
P = \frac{\nu T}{\lambda_T^d} \sum_n z^n b_n.
\label{eq:virial_definition}
\end{equation} 

\noindent Here $\nu$ is the spin degeneracy, which is $3$ for the one dimensional case and $2$ for the two dimensional case, and $d$ is the dimension. With Eq.~(\ref{eq:virial_definition}), we can identify the last term as the shift in the virial coefficient:

\begin{equation}
\delta b_3 = \frac{1}{6\pi}\int_{-\infty}^{\infty} d \epsilon \ e^{-\beta \epsilon} Im \left[ \frac{T_3(\epsilon-i \delta)}{-\epsilon + i \delta}\right].
\end{equation}

\noindent With our explicit expression for $T_3(\epsilon- i\delta)$, one can show that the imaginary part of $T_3(\epsilon - i \delta)$ has the form:

\begin{align}
Im[T_3(\epsilon - i \delta)] &= 2 \pi \sqrt{3} E_B \pi \delta(\epsilon + E_B) \nonumber \\
& \ +  \frac{2\pi^2 \sqrt{3}}{\ln^2\left(\frac{E_B}{\epsilon}\right) + \pi^2 }\theta(\epsilon),
\end{align}

\noindent which leads to the following expression for the virial coefficient:

\begin{equation}
\delta b_3 =  \frac{1}{\sqrt{3}}\left[e^{\beta E_B} - \int_0^{\infty}  \frac{d\epsilon}{\pi}\frac{e^{-\beta \epsilon}}{\epsilon}\frac{\pi}{\ln^2\left( \frac{E_B}{\epsilon}\right) + \pi^2}\right].
\label{eq:beth_uhlen}
\end{equation}

\noindent Eq.~(\ref{eq:beth_uhlen}) is no more than the famed Beth-Uhlenbeck formula \cite{BU}. Moreover, comparing to the two dimensional case \cite{Ordonez18b}, we can identify the following relation:

\begin{equation}
\delta b_3 = \frac{1}{\sqrt{3}} \delta b_2,
\label{eq:virial_coefficient_relations}
\end{equation}

\noindent where $\delta b_2$ is the shift in the second virial coefficient for the anomalous two-dimensional Fermi gas. This relationship was obtained previously in Ref.~\cite{Ordonez18a}

The presence of $E_B$ in the pressure will lead to a non-zero contact. The contact can be defined using the following relations \cite{Ordonez18a}:

\begin{align}
PL &= 2 \langle H \rangle + 2C_3 \nonumber \\
C_3 & = L E_B \frac{\partial P}{\partial E_B} \nonumber \\
&= \frac{g^2}{2\pi \sqrt{3}} \int dx \left\langle \psi_1^{\dagger}(x)\psi_2^{\dagger}(x)\psi_3^{\dagger}(x)\psi_3(x)\psi_2(x)\psi_1(x)\right\rangle.
\label{eq:contact_def}
\end{align}

From Eqns.~(\ref{eq:virial_definition}) and (\ref{eq:contact_def}), we can write down the definition of the contact as:

\begin{align}
\tilde{C_3} = \frac{C_3}{L} &= \frac{3 T}{\lambda_T} z^3 E_B\frac{\partial \delta b_3}{\partial E_B},
\end{align}

\noindent and similarly for two dimensions:

\begin{align}
\tilde{C_2} = \frac{C_2}{L} &= \frac{2 T}{\lambda_T^2} z^2 E_B\frac{\partial \delta b_2}{\partial E_B},
\end{align}

Using the relationship between the virial coefficients, Eq.~(\ref{eq:virial_coefficient_relations}), one can then show:

\begin{equation}
\tilde{C}_3 = \frac{\sqrt{3}}{2} z \lambda_T \tilde{C}_2,
\label{eq:Contact_relation}
\end{equation}

\noindent where $\tilde{C}_2 = C_2/L^2$ is the two dimensional contact density. The first factor comes from the fact that the spatial dimensions are different, and as a result, the dimensions of the contact will be different. The second factor is due to the relation between the coupling constants, Eq.~(\ref{eq:g_ratio}).

\noindent We can explicitly check Eq.~(\ref{eq:Contact_relation}) by evaluating the contact from Eq.~(\ref{eq:pressure}):

\begin{align}
\tilde{C}_3 = \frac{C_3}{L} &= \frac{z^3}{\lambda_T^3} \int_{-\infty}^{\infty} \frac{dx}{\pi} e^{-x} Im[T_3(x-i\delta, \beta E_B)] \nonumber \\
&= \frac{z^3}{\lambda_T^3} 2\pi\sqrt{3} \left[ \beta E_B e^{\beta E_B} \right. \nonumber \\
& \left. + \int_0^{\infty} dx e^{-x} \frac{1}{\log^2(\beta E_B/x) + \pi^2}  \right].
\label{eq:c3_exact}
\end{align}

\noindent This result is consistent with a perturbative calculation performed in Ref.~\cite{Pastukhov19} which evaluated the contact for the ground state at zero temperature. Eq.~(\ref{eq:c3_exact}) ought to be compared to the contact density of the anomalous two dimensional Fermi gas:

\begin{align}
\tilde{C}_2 &= \frac{C_2}{L^2} = \frac{z^2}{\lambda_T^2} 4\pi \left[ \beta E_B e^{\beta E_B} \right. \nonumber \\
& \left. + \int_0^{\infty} dx e^{-x} \frac{1}{\log^2(\beta E_B/x) + \pi^2}  \right].
\end{align}

\noindent  Upon comparison one obtains Eq.~(\ref{eq:Contact_relation}), the relationship between the contact densities.

\section{The Bulk Viscosity Spectral Function}
\label{sec:bulk_visc}

With the contact identified, we now turn to the bulk viscosity. The bulk viscosity can be defined as:

\begin{align}
\zeta(\omega) &= \frac{Im[\chi(\omega)]}{\omega} \nonumber \\
\chi(\omega) &= \frac{i}{Z L} \int_0^{\infty} dt e^{i(\omega+i \delta)t} Tr\left[e^{-\beta(H-\mu N)} [\Pi(t),\Pi(0)] \right],
\label{eq:bulk_def}
\end{align}

\noindent where $\Pi(t)$ is the spatially integrated stress-energy tensor. It is important to note that in one dimensional systems, the stress-energy tensor will satisfy

\begin{equation}
\Pi = P L = 2 \langle H \rangle + 2C_3.
\label{eq:pi}
\end{equation}

\noindent Substituting Eq.~(\ref{eq:pi}) into Eq.~(\ref{eq:bulk_def}), and noting that the thermal average of the commutator between $H$ and $C_3$ is defined to be zero when the system is in equilibrium, one obtains:

\begin{equation}
\chi(\omega) = 4\frac{i}{Z L} \int_0^{\infty} dt e^{i(\omega+i \delta)t} Tr\left[e^{-\beta(H-\mu N)} [C_3(t),C_3(0)] \right],
\label{eq:def_chi}
\end{equation}

Since the contact is a three-body operator, the first non vanishing term of $\chi(\omega)$ will be of order $O(z^3)$. We can then perform the trace over the three-body sector of the Hilbert space, and obtain an expression using Eqs.~(\ref{eq:prop_int}) and (\ref{eq:contact_def}). For explicit details on how to evaluate the trace, we refer the reader to Appendix \ref{app:correlator}, here we quote the final expression:

\begin{align}
\chi(\omega) = - 4z^3 \left( \frac{1}{2\pi \sqrt{3}}\right)^2  \frac{\sqrt{3}}{\lambda_T} &\int_{-\infty}^{\infty} \frac{d\epsilon}{\pi}\int_{-\infty}^{\infty} \frac{d\epsilon'}{\pi} \frac{e^{-\beta \epsilon}- e^{-\beta \epsilon'}}{\epsilon - \epsilon' + \omega + i \delta} \nonumber \\
&Im[T_3(\epsilon - i \delta)] Im[T_3(\epsilon' - i \delta)].
\label{eq:chi_final}
\end{align}

\noindent Substituting this into  Eq.~(\ref{eq:bulk_def}), we obtain the bulk viscosity:

\begin{align}
\zeta(\omega) &= 4z^3 \left( \frac{1}{2\pi \sqrt{3}}\right)^2  \frac{\sqrt{3}}{\lambda_T} \frac{1-e^{-\beta \omega}}{\omega} \nonumber \\
& \int_{-\infty}^{\infty} \frac{d\epsilon}{\pi}  e^{-\beta \epsilon} Im[T_3(\epsilon - i \delta)]Im[T_3(\epsilon +\omega - i \delta)].
\label{eq:bulk_viscosity_expression}
\end{align}

\begin{figure*}
\includegraphics[scale=0.8]{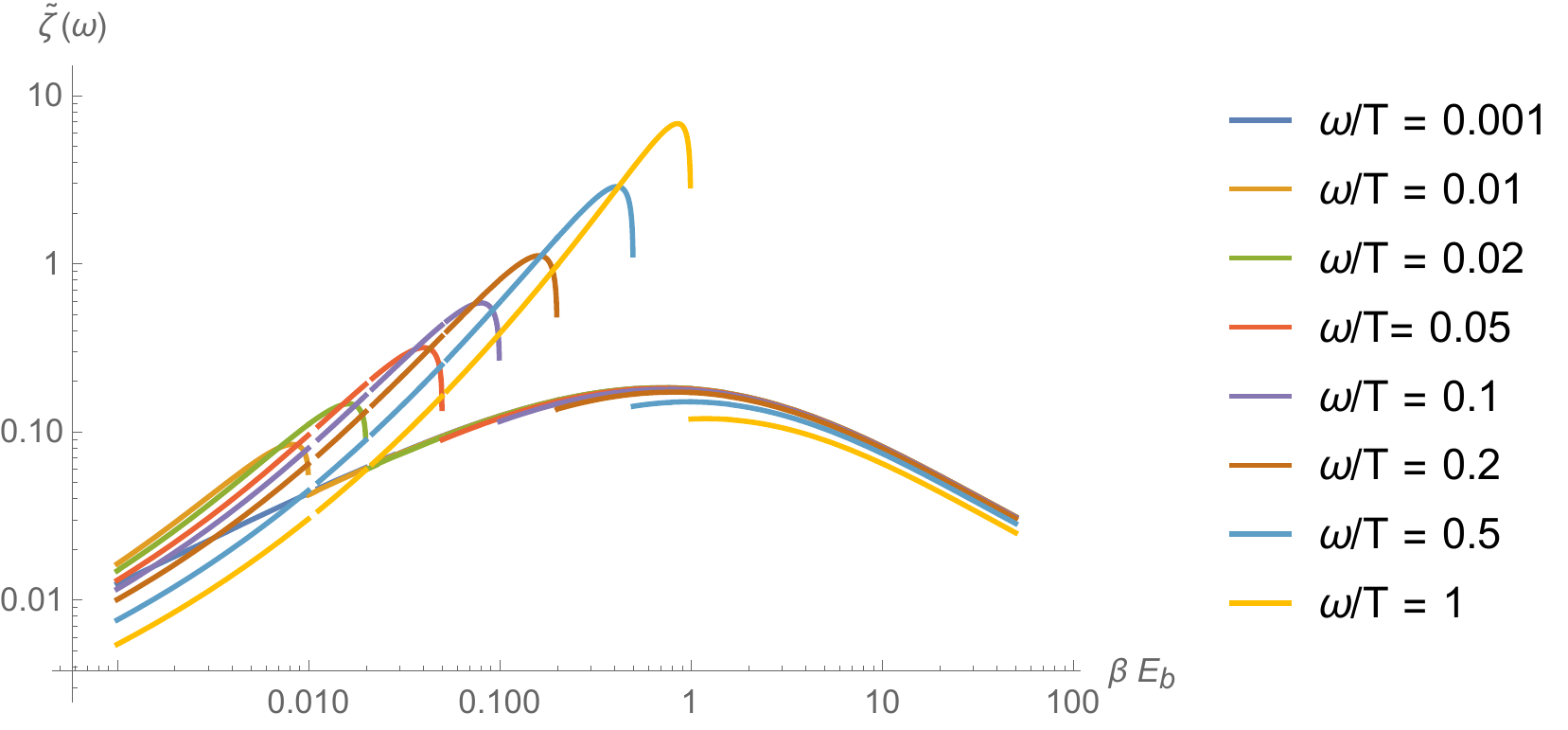}
\caption{The bulk viscosity as a function of binding energy, $E_B$. $\tilde{\zeta} = \zeta(\omega) (4z^3 /(2\pi\sqrt{3})^2)^{-1} \lambda_T/\sqrt{3}$. This structure is identical to the two dimensional bulk viscosity evaluated in Ref.~\cite{Nishida19}}
\label{fig:bulk_viscosity}
\end{figure*}

\noindent The bulk viscosity for various frequencies  as a function of $\beta E_B$ is shown in Fig.~(\ref{fig:bulk_viscosity}).

From Eq.~(\ref{eq:bulk_viscosity_expression}), one can show that the bulk viscosity is an even function of frequency, $\zeta(\omega) = \zeta(-\omega)$, and has the following large frequency limit:

\begin{equation}
\zeta(\omega\rightarrow \infty) = \frac{4\pi}{\log^2\left(\frac{\omega}{E_B}\right) \omega} \tilde{C}_3.
\end{equation}

\noindent One can also integrate Eq.~(\ref{eq:bulk_viscosity_expression}) to obtain the following sum rule:

\begin{align}
\int_{-\infty}^{\infty} \frac{d\omega}{\pi} \zeta(\omega) &= -\frac{4}{2\pi \sqrt{3}}E_B \frac{\partial}{\partial E_B} \tilde{C}_3.
\end{align}

\noindent Again as a comparison we note the bulk viscosity for 2D systems \cite{Nishida19} is given by:

\begin{align}
\zeta_2(\omega) &= z^2 \left( \frac{1}{4\pi }\right)^2  \frac{2}{\lambda_T^2} \frac{1-e^{-\beta \omega}}{\omega} \nonumber \\
& \int_{-\infty}^{\infty} \frac{d\epsilon}{\pi}  e^{-\beta \epsilon} Im[T_2(\epsilon - i \delta)]Im[T_2(\epsilon +\omega - i \delta)].
\end{align}

\noindent Or equivalently:

\begin{equation}
\zeta(\omega) = 4\frac{\sqrt{3}}{2} z \lambda_T \zeta_2(\omega).
\end{equation}

\section{Conclusions}
\label{sec:conc}

In this article we have explicitly confirmed that the thermodynamic properties of the anomalous one dimensional Fermi gas is directly related to that of the anomalous two dimensional Fermi gas. Thermodynamic properties like the virial coefficient, Tan's contact, and bulk viscosity, can all be related to one another, thanks to the mapping between the anomalous two-body physics in two dimensions and its three-body counterpart in one dimension. The mapping is summarized in Table ~\ref{tab:dictionary}.

This anomaly correspondence is an excellent tool in understanding the physics of anomalous systems at high temperatures, because the dominant contribution to the physics comes from the few-body sector. However, this mapping can not fully reproduce the entirety of the physics in both systems. The presence of an extra dimension will allow for the possibility of new phenomena which may have no counterpart for one dimensional systems. For example, a one dimensional system will not have a shear viscosity, but a two dimensional system will. 

In the future we will explore this relationship to see whether this mapping will be exact when many-body effects are important. To do this, it is necessary to examine the many-body properties of the anomalous one-dimensional Fermi gas, which is the subject of an upcoming work.

\begin{table}[]
\begin{tabular}{cc}
\\ \hline
\multicolumn{1}{|c|}{$g_3/g_2$}                         & \multicolumn{1}{c|}{$\sqrt{3}/2$}               \\ \hline
\multicolumn{1}{|c|}{$\delta b_3 / \delta b_2$}         & \multicolumn{1}{c|}{$1/\sqrt{3}$}                         \\ \hline
\multicolumn{1}{|c|}{$\tilde{C}_3 / \tilde{C}_2$}       & \multicolumn{1}{c|}{$\sqrt{3}/2 \  z \lambda_T$}   \\ \hline
\multicolumn{1}{|c|}{$\zeta(\omega) / \zeta_2(\omega)$} & \multicolumn{1}{c|}{$4 \sqrt{3}/2 \  z \lambda_T$} \\ \hline
\end{tabular}
\caption{The anomaly correspondence. Here we show the relation between various thermodynamic quantities for the anomalous one dimensional Fermi gas and the anomalous two dimensional Fermi gas. Here we note $g$ is the contact interaction, $\delta b$ is the virial coefficient, $C$ is Tan's contact, and $\zeta(\omega)$ is the bulk viscosity.}
\label{tab:dictionary}
\end{table}

\appendix

\numberwithin{equation}{section}
\renewcommand\theequation{\Alph{section}.\arabic{equation}}

\section{Explicit Forms for the Partition Function}
\label{app:z}

In this appendix we write down the explicit form of the partition function. We first note that the partition function can be written as:

\begin{align}
Z &= z Z_1 + z^2 \left(\frac{Z_1^2}{2} + \delta Z_2\right) \nonumber \\ 
&+ z^3 \left( \frac{Z_1^3}{3!} + Z_1 \delta Z_2 + \delta Z_3 + \left.Z_3\right|_{int}\right).
\end{align}

The various contributions to the partition function are given by:

\begin{align}
Z_1 &= \sum_k e^{-\beta k^2/2} \nonumber \\
\delta Z_2 &= -\frac{1}{2}\sum_k e^{-\beta k^2}, \nonumber \\
\delta Z_3 &= \frac{1}{3}\sum_k  e^{-3 \beta k^2/2}, \nonumber \\
\left. Z_3\right|_{int} &= \frac{1}{L^2}\sum_{Q}\sum_{p,q} \frac{2}{\sqrt{3}}e^{-\beta \frac{Q^2}{6}}  \int_{-\infty}^{\infty} \frac{d\epsilon}{\pi} \nonumber \\
& \left[ e^{-\beta \epsilon}  Im \left[ \frac{T_3(\epsilon - i \delta)}{(\epsilon - p^2 - q^2 )^2 }\right]\right].
\end{align}

\section{Calculation of the Retarded Contact-Contact Correlator}
\label{app:correlator}

In this section we write down an explicit form for the trace of the following quantity:

\begin{equation}
A = Tr_3\left[ e^{-\beta H} [C_3(t),C_3(0)] \right].
\end{equation}

\noindent This quantity is related to the retarded contact-contact correlator by:

\begin{equation}
\chi(\omega) = 4 \frac{iz^3}{L} \int_0^{\infty} dt e^{i (\omega+i \delta)t} A.
\label{eq:A_chi}
\end{equation}

It is important to note that $A$ can be rewritten as:

\begin{equation}
A = Tr_3\left[ e^{(-\beta+it)}H C e^{-i H t} - e^{(-\beta-it)H} C e^{i H t} \right]. 
\end{equation}

It is still possible to use the identities in Eqs.~(\ref{eq:prop}) and (\ref{eq:prop_int}) to express the evolution operators in terms of the propagator. One can then show that the trace of the contact-contact commutator is given by:

\begin{align}
A &= \left(\frac{g^2}{2\pi\sqrt{3} L^4}\right)^2 \sum_Q \sum_{p,q,p',q'} \sum_{k,l, k',l'} \int_{-\infty}^{\infty} \frac{d\epsilon}{\pi} \int_{-\infty}^{\infty} \frac{d\epsilon'}{\pi} \nonumber \\
& e^{-\beta \frac{Q^2}{6}} e^{i(\epsilon - \epsilon')t} (e^{-\beta \epsilon} - e^{-\beta \epsilon'}) \left(\frac{2}{\sqrt{3}}\right)^4\nonumber \\
&\left( Im\left[\frac{T_3(\epsilon - i \delta)}{(\epsilon - p^2 - q^2 - i \delta)(\epsilon - p'^2 - q'^2 - i \delta)} \right]  \right. \nonumber \\
& \left. Im\left[\frac{T_3(\epsilon' - i \delta)}{(\epsilon - k^2 - l^2 - i \delta)(\epsilon - k'^2 - l'^2 - i \delta)} \right]  \right).
\end{align}

\noindent Noting that:

\begin{equation}
\frac{g}{L^2} \frac{2}{\sqrt{3}}\sum_{p,q} \frac{1}{\epsilon - p^2 - q^2 - i \delta } \approx 1,
\end{equation}

\noindent when $\Lambda \rightarrow \infty$. One obtains the final result:

\begin{align}
A &= \left(\frac{1}{ 2\pi\sqrt{3}}\right)^2 \sum_Q e^{-\beta \frac{Q^2}{6}} \int_{-\infty}^{\infty} \frac{d\epsilon}{\pi} \int_{-\infty}^{\infty} \frac{d\epsilon'}{\pi} e^{i(\epsilon - \epsilon')t}\nonumber \\
& (e^{-\beta \epsilon} - e^{-\beta \epsilon'})Im\left[ T_3(\epsilon - i \delta)\right]Im\left[ T_3(\epsilon' - i \delta)\right].
\label{eq:A_final}
\end{align}

\noindent Substituting Eq.~(\ref{eq:A_final}) into Eq.~(\ref{eq:A_chi}) and performing the integrations over the center of mass momentum and time, one obtains Eq.~(\ref{eq:chi_final}).

\end{document}